\documentclass[prc,aps,floatfix,showpacs]{revtex4}
\usepackage{graphicx}
\usepackage{amssymb}
\usepackage{amsmath}
\usepackage{bm}

\begin{document}
\title{Coulomb energy contribution to the excitation energy in $^{229}$Th
 and enhanced effect  of $\alpha$ variation}
\author{V.V. Flambaum$^{1,2}$, N. Auerbach$^3$, and V.F. Dmitriev$^{1,3,4}$, }

\affiliation{$^1$
 School of Physics, The University of New South Wales, Sydney NSW
2052, Australia}
 \affiliation{$^2$ 
Perimeter Institute, 31 Caroline St.N, Waterloo, ON Canada N2L 2Y5}
\affiliation{$^3$
 School of Physics and Astronomy, Tel Aviv University, Tel Aviv, 69978, Israel
 }
\affiliation{$^4$
Budker Institute of Nuclear Physics, 630090, Novosibirsk-90,
Russia
}
\begin{abstract}
We calculated the contribution of Coulomb energy to the spacing between the
 ground and first excited state of $^{229}$Th nucleus as
 a function of the deformation parameter $\delta$. We show that despite the
 fact that the odd particle is a neutron, the change in Coulomb energy
 $\Delta U_C $ between these two states can reach several hundreds keV.
 This means that the effect of the variation of the
fine structure constant  $\alpha=e^2/\hbar c$ may be  enhanced 
 $\Delta U_C/E \sim  10^4$ times in the  $E=$7.6 eV  
``nuclear clock'' transition between the ground and first excited states 
in the $^{229}$Th nucleus.
\end{abstract}
\maketitle
\section{Introduction}
Unification  theories applied to cosmology suggest a possibility
of variation of the fundamental constants in
the expanding Universe (see e.g. review \cite{Uzan}).
There are  hints of variation of
 $\alpha$ and $m_{q,e}/\Lambda_{QCD}$
in quasar absorption spectra, Big Bang nucleosynthesis and Oklo
natural nuclear reactor data
(see \cite{var} and references therein).
 Here  $\Lambda_{QCD}$ is the quantum chromodynamics
 (QCD) scale,  and $m_{q}$ and $m_{e}$ are the quark and electron masses.
However, the majority of publications report only limits on  possible variations
 (see e.g. \cite{Uzan,F2007,karshenboim,Kozlov 2007,Murphy 2008}).
 A very sensitive method to study the
 variation in a laboratory
 consists of the comparison of different optical and microwave atomic clocks
(see e.g. measurements in
 \cite{prestage,Marion 2003,Bize 2005,Peik 2004,Bize 2003,
Fischer 2004,Peik 2005,Blatt,Rosenband}).
An  enhancement of the relative effect of $\alpha$ variation can be obtained
in a transition between almost degenerate levels in Dy atom \cite{dzuba1999}.
These levels move in opposite directions if  $\alpha$ varies.
 An experiment is currently underway to place limits on
$\alpha$ variation using this transition \cite{budker}, but
unfortunately one of the levels has  quite a large linewidth
and this limits the accuracy. An enhancement of 1-5 orders exists
in narrow microwave molecular transitions \cite{mol}.
Some atomic transitions with enhanced sensitivity are listed in
Ref.~\cite{at}.
In  Ref. \cite{th7} it was suggested that there may
be a five orders of magintude enhancement of the variation effects
in the low-energy transition between the
ground and the first excited states in the $^{229}$Th nucleus.
The exsitence of the enhancement was confirmed in \cite{He}.
This transition in  $^{229}$Th  was suggested as a possible nuclear
 clock in Ref.~\cite{th8}.
Indeed, the transition is very narrow.  
The width of the excited state is estimated to be about 
$10^{-4}$ Hz~\cite{th2} (the experimental limits on the width are given
in \cite{th3}). 
The latest measurement of the transition energy~\cite{Beck} gives
$7.6 \pm 0.5$ eV, compared to earlier values of $5.5\pm 1$ eV \cite{th6} 
and $3.5\pm 1$ eV \cite{th1}.
This makes $^{229}$Th a possible reference for an
 optical clock of very high accuracy, and opens a new possibility
for a laboratory search for the variation of the fundamental constants.
Large  interest in experiments exploring this possibility was expressed,
for example, in E. Hudson and E. Peik talks at  special workshop
 (14-18 July 2008, Perimeter Institute)  devoted to the variation 
of the fundamental constants
 and privite communications
of E. Peik, P. Beiersdorfer, Zheng-Tian Lu,  D. DeMille, D. Habs
 and J. Torgerson. This motivated us to perform an independent
calculation of the sensitivity of the transition frequency $\nu$ to
 the variation of $\alpha$ based on the following relation  suggested in 
 Ref.  \cite{Hayes}:
\begin{equation} \label{eq0}
h \delta \nu = \Delta U_C \frac{\delta \alpha}{\alpha}
\end{equation}
where  $\Delta U_C$ is the contribution of the Coulomb interaction to the
 energy spacing between these two levels.

\section{Shift of nuclear Coulomb energy}
\label{sec:cce}
The calculation of the difference in total Coulomb energy between ground and excited nuclear states can hardly be done within existing nuclear models with the required accuracy.
Instead, the calculation of the correction to the Coulomb energy when one proton or one neutron is added into a single particle orbital $|\nu\rangle $ is a much simpler problem.

The ground state of  $^{229}$Th nucleus is $[Nn_z\Lambda\, J^P]=[633\,5/2^+]$;
i.e. the deformed oscillator quantum numbers are $N=6$, $n_z=3$, the
 projection of the valence neutron orbital angular momentum on the
nuclear symmetry axis (internal  z-axis)
 is $\Lambda=3$, the spin projection $\Sigma=-1/2$, and the
 total angular momentum and the total
 angular momentum projection are $J=\Omega=\Lambda+\Sigma=5/2$.
The excited state is $[Nn_z\Lambda\, J^P]=[631\,3/2^+]$; i.e.
it has the same  $N=6$ and  $n_z=3$. The values
$\Lambda=1$, $\Sigma=1/2$ and  $J=\Omega=3/2$ are different.
With relation to the variation of alpha, we shall discuss below the
contribution
of the Coulomb interaction to the energy spacing between these two levels.

We start from an even $^{228}$Th nucleus and  calculate the shift of
nuclear Coulomb energy when an odd particle is added into single particle state $|\nu\rangle $. In an even nucleus the Coulomb energy can be presented as
\begin{equation} \label{eq1}
U_C=\frac{1}{2}\int d^3r\,d^3r' \frac{\rho_c({\bm r})\rho_c({\bm r}')}
{|{\bm r}-{\bm r}'|},
\end{equation}
where $\rho_c({\bm r})$ is the charge density. For uniformly charged sphere we obtain the well known expression
$$
U_C=\frac{3}{5}\frac{Z^2e^2}{R},
$$
which gives for Z=90, and A=228, the estimate, $U_C\sim$1 GeV. The above equation can also be used to estimate the change in the Coulomb energy when one neutron is added. An addition of the neutron changes A, A$\rightarrow$A+1, therefore
$$
\Delta U_C = -\frac{U_C}{3A},
$$
that gives for $\Delta U_C$ the value of the order of -1 MeV. This estimate gives some average macroscopic value for $\Delta U_C$ because in reality it depends on the state $|\nu\rangle $ where we put the neutron.

 Adding an odd particle to the state
 $|\nu\rangle $ we obtain a change in the charge density
\begin{equation} \label{eq2}
\langle\nu|\rho_c({\bm r})|\nu\rangle = \rho_c({\bm r})
+\delta\rho^{\nu\nu}({\bm r}),
\end{equation}
where $\rho_c({\bm r})$ is the charge density of a neighbor even nucleus.
For the change in the Coulomb energy Eq.(\ref{eq1}) we obtain
\begin{equation} \label{eq3}
\Delta U_C^\nu = \int d^3r U_c({\bm r})\delta\rho^{\nu\nu}({\bm r}),
\end{equation}
where $U_c({\bm r})$ is the single particle Coulomb potential
\begin{equation} \label{eq4}
U_c({\bm r})=\int d^3r'\frac{\rho_c({\bm r})}{|{\bm r}-{\bm r}'|}.
\end{equation}
Although $^{228}$Th is a deformed nucleus, for the estimates we shall neglect a small quadrupole component in the charge density $\rho_c({\bm r})$. Effects of deformation will be discussed below. In approximation of uniformly charged sphere
\begin{equation} \label{eq5}
U_c({r})=\left\lbrace  \begin{array}{l}\frac{Ze^2}{R}(\frac{3}{2}-\frac{r^2}{2R^2})\;\; r\leq R, \\
  \\
\frac{Ze^2}{r} \;\;  r > R,\end{array} \right.
\end{equation}
where $R$ is a Coulomb radius of $^{228}$Th, $R=1.2 A^{1/3}$fm. We use this very simple form in order to obtain an
estimate for the change in the Coulomb energy when going from one
nuclear state to another state, and not in the absolute value.

The ground state of the nucleus $^{229}$Th is polarized due to the addition of the nucleon to the $^{228}$Th core. The added nucleon introduces an admixture of the monopole state (which is a radial excitation) \cite{au71,au74} and the nuclear state is now:
\begin{equation} \label{eq12}
|0'\nu\rangle =\sqrt{1-\epsilon^2}|0^+;\phi_\nu\rangle + \epsilon |M;\phi_\nu\rangle,
\end{equation}
where $|0^+\rangle$ and $|M\rangle$ are the ground state and the isovector monopole in $^{228}$Th and $\phi_\nu$ is the wave function of the added nucleon. With
\begin{equation} \label{eq13}
\epsilon = \frac{\langle 0^+;\phi_\nu|F_N|M;\phi_\nu\rangle}{E_0-E_M},
\end{equation}
where $F_N$ is a two-body nuclear interaction.

Now we evaluate the Coulomb energy of this new polarized state $|0'\nu\rangle$. The correction is:
\begin{equation} \label{eq14}
\Delta U^\nu_C=2\epsilon\langle 0^+;\phi_\nu|\frac{1}{2}\sum_{i\neq j}V_C({\bm r}_i-{\bm r}_j)|M;\phi_\nu\rangle,
\end{equation}
$V_C(r)$ is the Coulomb potential. This correction can be written in the form:
\begin{equation} \label{eq15}
\Delta U^\nu_C=2\epsilon\frac{Z}{A}\int \delta\rho(r)U_c(r)d^3r.
\end{equation}
$\delta\rho(r)$ is the transition density between the ground state and the monopole:
$$
\delta\rho =\langle M|\psi^\dagger({\bm r})t_z\psi({\bm r})|0^+\rangle.
$$
The Z/A factor takes into account the fact that only protons interact via the Coulomb.

The strong interaction matrix element we evaluate using a simple interaction:
\begin{equation} \label{eq16}
F_N=F_0\delta ({\bm r}_1-{\bm r}_2).
\end{equation}
Then:
\begin{equation} \label{eq17}
\langle 0^+;\phi_\nu|F_N|M;\phi_\nu\rangle = F_0\int \delta\rho(r)|\phi_\nu({\bm r})|^2d^3r.
\end{equation}

Finally:
\begin{equation} \label{eq18}
\Delta U^\nu_C=\frac{2Z/AF_0\int \delta\rho(r)U_C(r)d^3r\int \delta\rho(r)|\phi_\nu({\bm r})|^2d^3r}{E_0-E_M}.
\end{equation}
The result obviously depends on $\phi_\nu$.

One form used for the monopole transitions is:
\begin{equation} \label{eq19}
\delta\rho=C(3\rho+r\frac{d\rho}{dr})\equiv C\frac{1}{r^2}\frac{d(r^3\rho)}{dr},
\end{equation}
where $C$ is a constant and $\rho$ is the ground state density. This form
for the transition density corresponds to a change obtained by uniformly
expanding the nucleus, but also can be derived from a sum
rule \cite{au71,au74}. This transition density has a node. Because of the node in $\delta\rho$ the result for the polarization correction will be sensitive to the spatial form of the wave function $\phi_\nu$, and even may change sign for different single-particle states.

To obtain a quantitative estimate let us switch to the particle-hole basis $|ph\rangle$ instead of the single monopole resonance $|M\rangle$.
The correction to the charge density introduced in Eq.(\ref{eq2}) can be presented as follows:
\begin{equation} \label{eq6}
\delta\rho^{\nu\nu}({\bm r})= \sum_{\lambda,\lambda^\prime} \delta\rho^{\nu\nu}_{\lambda\lambda^\prime}\psi^\dagger_{\lambda^\prime} (\bm r)\psi_\lambda (\bm r),
\end{equation}
where $\psi_\lambda (\bm r)$ is the single-particle wave function, and $\delta\rho^{\nu\nu}_{\lambda\lambda^\prime}$ is the correction to the proton density matrix of a neighbohr even nucleus, due to the odd nucleon added into the state $|\nu \rangle$. 
\begin{equation} \label{eq77}
\langle \nu|a_{\lambda^\prime}^{p\dagger}a^p_\lambda |\nu\rangle \approx \langle 0|a_{\lambda^\prime}^{p\dagger}a^p_\lambda |0\rangle +\delta\rho^{\nu\nu}_{\lambda\lambda^\prime}.
\end{equation}
Here $\langle 0|a_{\lambda^\prime}^{p\dagger}a^p_\lambda |0\rangle $ is the proton density matrix of the neighbohr even nucleus.

Substituting Eq.(\ref{eq6}) into Eq.(\ref{eq3}) we obtain
\begin{equation} \label{eq7}
\Delta U_C^\nu =\sum_{\lambda,\lambda^\prime} \left( U_c\right)_{\lambda^\prime\lambda}\delta\rho^{\nu\nu}_{\lambda\lambda^\prime},
\end{equation}
where $\left( U_c\right)_{\lambda^\prime\lambda}$ is a matrix element of the single-particle Coulomb potential Eq.(\ref{eq5}).

The equation for $\delta\rho^{\nu\nu}_{\lambda\lambda^\prime}$ can be found for example in \cite{mig}.
It states
\begin{equation} \label{eq8}
\delta\rho^{\nu\nu}_{\lambda\lambda^\prime}= \delta_{\nu\lambda}\delta_{\nu\lambda^\prime}+\dfrac{n_\lambda-n_{\lambda^\prime}}
{\epsilon_\lambda-\epsilon_{\lambda^\prime}}\sum_{\lambda_1\lambda_2}
\langle\lambda\lambda_1|F_N|\lambda_2\lambda^\prime\rangle \delta\rho^{\nu\nu}_{\lambda_2\lambda_1}.
\end{equation}
Here $\langle\lambda\lambda_1|F_N|\lambda_2\lambda^\prime\rangle$ is the matrix element of effective residual interaction between quasi-particles. The first term in r.h.s. of Eq.(\ref{eq8}) exists only for an odd proton. If the odd particle is a neutron, this term is absent. The second term is the polarization correction which exists both for an odd proton and an odd neutron.

In the first order in proton neutron residual interaction we obtain for the correction to the proton density matrix
\begin{equation} \label{eq9}
\delta\rho^{\nu\nu}_{\lambda\lambda^\prime}=\dfrac{n_\lambda-n_{\lambda^\prime}}
{\epsilon_\lambda-\epsilon_{\lambda^\prime}}
\langle\lambda\nu|F^{pn}_N|\nu\lambda^\prime\rangle,
\end{equation}
where the state $|\nu \rangle $ refers to a neutron. The correction to Coulomb energy becomes
\begin{equation} \label{eq10}
\Delta U_C^\nu =\sum_{\lambda,\lambda^\prime} \left( U_c\right)_{\lambda^\prime\lambda}\dfrac{n_\lambda-n_{\lambda^\prime}}
{\epsilon_\lambda-\epsilon_{\lambda^\prime}}
\langle\lambda\nu|F^{pn}_N|\nu\lambda^\prime\rangle.
\end{equation}
Eq.(\ref{eq10}) can be presented in the following way,
\begin{equation} \label{eq11}
\Delta U_C^\nu = \int d^3r \delta U_C(r) |\psi_\nu ({\bm r})|^2.
\end{equation}
Here we introduced the correction to potential energy $\delta U_C(r)$ that describes the reaction of the proton core to the presence of an additional neutron 
\begin{equation} \label{111}
\delta U_C(r) = F_0\sum_{\lambda,\lambda^\prime} \left( U_c\right)_{\lambda^\prime\lambda}\dfrac{n_\lambda-n_{\lambda^\prime}}
{\epsilon_\lambda-\epsilon_{\lambda^\prime}} \psi_\lambda^\dagger ({\bm r})\psi_{\lambda^\prime}({\bm r}).
\end{equation}
It follows from Eqs.(\ref{eq10},\ref{111}) that the correction $\delta U_C(r)$ does not depend on the neutron state. However, its mean value given by Eq.(\ref{eq11}) does depend on the particular state $|\nu\rangle$. The radial dependence of $\delta U_C(r)$ is shown in Fig.1.  
\begin{figure}[h]
\includegraphics {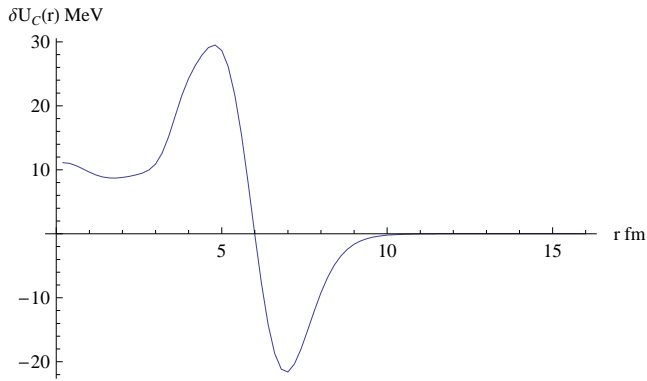}
\caption{Radial dependence of the correction to the Coulomb potential energy}
\end{figure}
This radial dependence shown is typical for a double charged layer. Due to small compressibility of a nucleus the change of a charge density can take place on nuclear surface only, and in good approximation it can be presented as a double charged layer. The mean value of the correction $\delta U_C(r)$ over nuclear volume is negative and equal to -440 keV. Due to oscillating behavior, the expectation value of $\delta U_C(r)$ will obviously be state dependent.

The calculations for $^{229}$Th were performed using Nilsson wave functions \cite{chi} for the valence neutron in order to account for deformation.  In $^{229}$Th the ground state has the quantum numbers [633 5/2+] while the excited state has [631 3/2+]. Both of them belong to N=6 shell where the possible angular momenta in the Nilsson orbitals are L=2, 4, 6.  The shift of Coulomb energy calculated due to add of valence neutron into these states is shown in Table 1.
\begin{table}[h]
\caption{Shifts of the Coulomb energy due to odd neutron for N=6 shell orbitals.}
\begin{tabular}{l|c|c|c|c|c|c}
 & $3d_{3/2}$& $3d_{5/2}$& $2g_{7/2}$& $2g_{9/2}$&$1i_{11/2}$&$1i_{13/2}$ \\
 \hline
 $\Delta U_C$ (MeV) & -0.741& -0.926 & -0.848 & -1.381& +0.158&+1.305 \\
\end{tabular}
\end{table}
Note, that the shift for $1i_{13/2}$ and $1i_{11/2}$ becomes positive.
The radial wave function for L=6 has a sharp peak near nuclear surface where $\delta U_C(r)$ is positive. The state $1i_{11/2}$ has smaller binding energy and its peak, compared to $1i_{13/2}$ state, is shifted to larger r where $\delta U_C(r)$ becomes negative. This leads to a smaller value of the shift for $1i_{11/2}$ state.

The difference between the first excited state with quantum numbers [631 3/2] and the ground state [633 5/2] in $^{229}$Th is extremely small ($<$ 10 eV) and can not be reproduced in Nilsson model at reasonable deformation $\delta \backsimeq 0.2$. With the standard parameters of the Nilsson model \cite{chi} these two levels are crossed at the deformation $\delta \leq 0.1$.  Therefore, we calculated the difference in Coulomb energies $\Delta U_C^{3/2}-\Delta U_C^{5/2}$ at different deformations, as a function of the parameter $\eta=\delta/\kappa$, where $\kappa=0.05$ is the spin-orbit constant. The result is shown in Fig.2 for $0\leq \eta \leq 6$. This interval corresponds to deformations $0 \leq \delta \leq 0.3$.
\begin{figure}[h]
\includegraphics {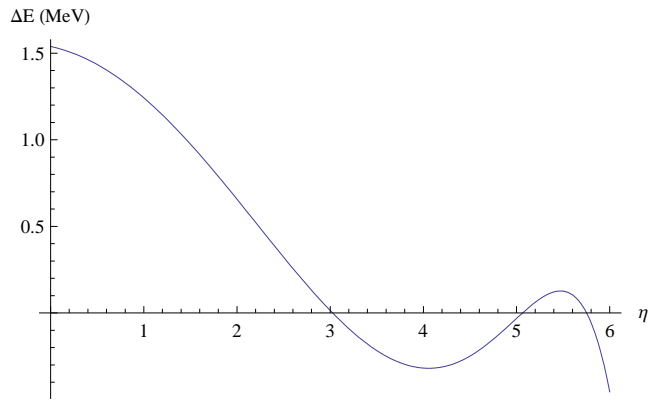}
\caption{Difference in Coulomb energies between the first excited and the ground states in $^{229}$Th as a function of deformation.}
\end{figure}
Note that the difference becomes negative at the deformations $0.15 \leq \delta \leq 0.25$. However, the position of this interval depends on the parameters of the Nilsson model, while the fact that the difference changes sign at some deformation remains in all sets of the parameters. The value of $\Delta U_C $ as a function of deformation changes from 1.5 MeV at zero deformation down to -0.5 MeV at $\delta = 0.3$. In our approach we can hardly give a better estimate, due to simplified treatment of deformation. However, very small value of the Coulomb energy shift seems unprobable.
Note that near the Nilsson level crossing
the difference of the Coulomb energies is about 1 MeV.
 The shift of the order from several tens keV to several hundreds keV provides 
an enhancement of sensitivity to changes of $\alpha$, as it was mentioned in
 \cite{th7}. Indeed, the frequency shift from  Eq.(\ref{eq0}) may be presented
 in the following form  
\begin{equation} \label{eq20}
\delta \nu = 
 \frac{\Delta U_C}{100 KeV} \frac{\delta \alpha}{\alpha} 2.42\times 10^{19} Hz
\end{equation}
This shift is four to five orders of magnitude larger than the shift 
which was studied in optical atomic clocks
 \cite{Peik 2004,Fischer 2004,Peik 2005,Blatt,Rosenband}). An additional
 enhancement may be due to the very small width of this transition
($10^{-4}$ Hz~\cite{th2}). It is also instructive to present the relative
enhancement:
\begin{equation} \label{eq21}
\frac{\delta \nu}{\nu} = 
 \frac{\Delta U_C}{100 KeV}\frac{100 KeV}{7.6 eV}\frac{\delta \alpha}{\alpha}=
 1.3 \times 10^{4}\frac{\Delta U_C}{100 KeV}\frac{\delta \alpha}{\alpha}
\end{equation}
These estimates confirm conclusion of Ref. \cite{th7} that experiments
with $^{229}$Th have potential to improve sensitivity of laboratory
measurements of the variation of the fundamental constants up to 8 orders
of magnitude.

\acknowledgments

N.A. thanks Byron Jennings for the hospitality at TRIUMF, V.F. is grateful to
 Perimeter
Institute for the hospitality and support.
This work was supported in part by the US-Israel Binational Science
Foundation and Australian Research Council

\end{document}